\documentclass[prd,twocolumn,reprint,preprintnumbers,nofootinbib,superscriptaddress,longbibliography]{revtex4-1}

\usepackage{soul}
\usepackage{verbatim}
\usepackage{booktabs}
\usepackage{siunitx}
\usepackage{hyperref}
\usepackage{tabularx}
\usepackage{multirow}
\usepackage{tikz}
\usetikzlibrary{tikzmark, calc}
\usepackage{amsmath}
\usepackage{amsthm}
\usepackage{amssymb}
\usepackage{graphicx}
\usepackage{ulem}
\makeatletter
\makeatletter \renewcommand{\@dotsep}{10000} \makeatother
\usepackage[utf8]{inputenc}
\usepackage{lmodern}
\usepackage{hyperref}
\hypersetup{
	unicode,
	colorlinks,
    breaklinks=true,	urlcolor=cyan, 
    linkcolor=black, 
	pdfauthor={Author One, Author Two, Author Three},
	pdftitle={A simple article template},
	pdfsubject={A simple article template},
	pdfkeywords={article, template, simple},
	pdfproducer={LaTeX},
	pdfcreator={pdflatex}
}
\usepackage[T1]{fontenc} 
\usepackage{amsmath}
\usepackage{graphicx, color}
\usepackage{physics}
\usepackage{url}

\usepackage{bm}
\usepackage{xcolor}
\usepackage{adjustbox}
\usepackage{tikz}
\usetikzlibrary{quantikz2, positioning, calc}

\definecolor{group1}{RGB}{244, 96, 54}
\definecolor{group2}{RGB}{8, 124, 167}
\definecolor{gate}{RGB}{198, 216, 211}
\tikzset{
operator/.append style={fill=gate!20},
meter/.append style={fill=gate!20},
}

\newcommand{\be}{\begin{eqnarray}}
\newcommand{\ee}{\end{eqnarray}}
\def\be{\begin{equation}}
\def\ee{\end{equation}}
\def\bea{\begin{eqnarray}}
\def\eea{\end{eqnarray}}

\newcommand{\gsim}{\;\raisebox{-0.9ex}{$\textstyle\stackrel{\textstyle >}{\sim}$}\;}
\newcommand{\lsim}{\;\raisebox{-0.9ex}{$\textstyle\stackrel{\textstyle<}{\sim}$}\;}
\def\lsim{\raise0.3ex\hbox{$\;<$\kern-0.75em\raise-1.1ex\hbox{$\sim\;$}}}
\def\gsim{\raise0.3ex\hbox{$\;>$\kern-0.75em\raise-1.1ex\hbox{$\sim\;$}}}

\usepackage{graphics}
\usepackage{epsfig}
\usepackage{slashed}
\usepackage[utf8]{inputenc}

\usepackage{multirow}
 \usepackage{pstricks}
\usepackage{dcolumn}

\theoremstyle{plain}

\theoremstyle{definition}

\makeatletter
\newcommand{\fmslash}[2][0mu]{%
  \mathchoice
    {\fmsl@sh\displaystyle{#1}{#2}}%
    {\fmsl@sh\textstyle{#1}{#2}}%
    {\fmsl@sh\scriptstyle{#1}{#2}}%
    {\fmsl@sh\scriptscriptstyle{#1}{#2}}}
\newcommand{\fmsl@sh}[3]{%
  \m@th\ooalign{$\hfil#1\mkern#2/\hfil$\crcr$#1#3$}}
\makeatother

\begin{document}

\author{Zhongtian Dong}
\email{cdong@ku.edu}
\affiliation{Department of Physics and Astronomy, University of Kansas, Lawrence, KS 66045, USA}

\author{Doojin Kim}
\email{doojin.kim@usd.edu}
\affiliation{Department of Physics, University of South Dakota, Vermillion, SD 57069, USA}

\author{Kyoungchul Kong}
\email{kckong@ku.edu}
\affiliation{Department of Physics and Astronomy, University of Kansas, Lawrence, KS 66045, USA}

\author{Myeonghun Park}
\email{parc.seoultech@seoultech.ac.kr}
\affiliation{Center for Theoretical Physics of the Universe,
Institute for Basic Science, Daejeon 34126, South Korea}
\affiliation{Institute of Convergent Fundamental Studies, Seoultech, Seoul 01811, South Korea}
\affiliation{School of Natural Sciences, Seoultech,  Seoul 01811, South Korea}

\author{Miguel A. Soto Alcaraz}
\email{msoto1@ku.edu}
\affiliation{Department of Physics and Astronomy, University of Kansas, Lawrence, KS 66045, USA}

\title{Quantum Sensing Radiative Decays of Neutrinos and Dark Matter Particles}

\begin{abstract} 
We explore a novel strategy for detecting the radiative decay of very weakly interacting particles by leveraging the extreme sensitivity of quantum devices, such as superconducting transmon qubits and trapped ion systems, to faint electromagnetic signals.
By modeling the effective electric field induced by the decay photons, we evaluate the response of quantum sensors across two particle physics scenarios: the cosmic neutrino background and two-component dark matter. We assess the discovery potential of these devices and outline the parameter space accessible under current experimental capabilities. Our analysis demonstrates that quantum sensors can probe radiative decays of dark matter candidates using existing technology, while probing neutrino magnetic moments beyond current limits will require scalable quantum architectures with collective enhancement.
\end{abstract}

\maketitle

\section{Introduction}

The existence of dark matter (DM) is one of the most compelling arguments for new physics beyond the standard model (SM). 
A myriad of experimental searches, tailored to different mass ranges and interaction strengths, have been performed for several decades and continue to advance. 
Yet its true nature remains elusive, and a wide range of theoretical scenarios have been proposed. In particular, the possible mass scale of DM is remarkably broad and poorly constrained, spanning from ultra-light particles as light as $10^{-24}$ eV to massive astrophysical objects such as primordial black holes with masses exceeding millions of solar masses 
\cite{Cooley:2022ufh,Bozorgnia:2024pwk}. 

Similarly, the Cosmic Neutrino Background (C$\nu$B) is a fundamental prediction of Big Bang cosmology, representing relic neutrinos that decoupled from the primordial plasma roughly one second after the Big Bang \cite{Giunti:2014ixa,Giunti:2015gga,Broggini:2012df}. Yet, both DM and the C$\nu$B remain undetected by direct means, largely due to their extremely weak interactions with ordinary matter.

In recent years, numerous proposals have been put forward to detect DM candidates using quantum devices, including searches for hidden photon DM \cite{Chigusa:2025rqs,Chen:2022quj,He:2024sfi}, axions \cite{Chen:2024aya}, and light DM \cite{Kim:2020coy, Ito:2023zhp, Linehan:2024btp, Chao:2024owf,Badurina:2025idj}.
To enhance signal detection and suppress noise, several algorithmic approaches have also been explored \cite{Chen:2023swh, Shu:2024nmc}. A potential opportunity for directional detection has also been explored in Ref. \cite{Fukuda:2025zcf}. 
On the experimental front, significant progress has been made, such as the development of flux-tunable cavities \cite{Zhao:2025thg}, transmon qubit modeling \cite{Moretti:2024xel, Labranca:2025bvc}, and scalable architectures for dark photon searches \cite{Kang:2025kaf}. For a broader overview of quantum sensors in high-energy physics, we refer the reader to Ref. \cite{Chou:2023hcc}. 

Along these lines, the DarQ collaboration recently started taking data and presented first results, excluding the dark photon parameter region for a dark photon mass around 36.1 $\mu$eV with a peak sensitivity of  $\sim 10^{-12}$ for the kinetic mixing parameter, surpassing the existing cosmological bounds \cite{Nakazono:2025tak}. 

While most existing studies focus on (bosonic) axion and hidden photon DM, 
in this paper, we explore the use of quantum devices\footnote{We henceforth use the terms, quantum device and qubit device, interchangeably.} to detect the radiative decays of extremely weakly interacting particles to SM photons. In particular, we consider a two-component DM scenario as well as the C$\nu$B as test cases, and explore the potential of leveraging qubit sensors as effective single-photon detectors while treating the associated photon field in a semi-classical manner. 
By modeling the effective electric field produced by decay photons and their interaction with qubit sensors, we evaluate their discovery potential and map out the parameter space accessible with near-future experimental capabilities. Our study suggests that current quantum technologies can effectively probe radiative DM decays, whereas pushing the sensitivity to neutrino magnetic moments beyond present bounds will require more scalable quantum systems with collective enhancement. We also investigate potential signal enhancement using symmetric Dicke states.

This paper is structured as follows. 
Section \ref{sec:methods} introduces the theoretical framework. Section \ref{sec:appA} discusses the justification for the sinusoidal approximation. Section \ref{sec:analysis} presents the measurement protocol, while Section \ref{sec:dicke} discusses the potential enhancement from Dicke states. Section \ref{sec:sensitivity} applies the framework to dark matter and the C$\nu$B, and Section \ref{sec:conclusion} concludes.

\section{Event rates}
\label{sec:methods}

\subsection{Effective electric fields from particle decays}
\label{sec:Eeff}

We consider physics scenarios where the electromagnetic field induced by a photon from the particle decay causes the transition of quantum devices from the ground state to an excited state. 
Although we refer to a ``single photon'' for concreteness, 
the decay photons considered here are not assumed to constitute a coherent classical electromagnetic wave.  Instead, we use a response-level parametrization motivated by the quantum-optics absorption formula discussed in Section~\ref{sec:appA}. Namely, the near-resonant local field correlator seen by the sensor is encoded in an effective amplitude $\bar E^{(\rm eff)}$ and a carrier frequency $E_\gamma$, 
\begin{equation}
    \vec E_{(\rm eff)}(t) = \bar E_{\rm (eff)} \vec n_E \sin (E_\gamma t) \, , \label{eq:efffield}
\end{equation}
where $\vec n_E$ is the unit vector representing the direction of the effective electric field $\vec E_{(\rm eff)}$. 
The notation in Eq.~\eqref{eq:efffield} is therefore understood as a compact way to reproduce the near-resonant two-level detector response, not a statement that the decay-photon state has a nonzero coherent one-point field amplitude.
With this convention, we will see that, in Section~\ref{sec:appA}, the resulting absorption probability takes the same mathematical form as the resonant Rabi response of a two-level system, with an effective coupling proportional to $\bar E^{(\rm eff)}$. 
The absolute phase can be absorbed into a choice of time origin, and the transition probability depends primarily on the drive intensity.

As a concrete example, let us consider the decay of the heavier species $X_2$ into the lighter species $X_1$ and a photon:
\begin{equation}
    X_2 \to X_1 +\gamma, 
\end{equation}
for which the photon energy is given by 
\begin{equation}
    E_\gamma = \frac{m_2^2 - m_1^2}{2 m_2} = \frac{\Delta m^2}{2 m_2} \label{eq:Egamma}
\end{equation}
in the rest frame of $X_2$. Here $\Delta m^2 \equiv m_2^2 - m_1^2$ denotes the squared-mass difference.

We assume that the particle $X_2$ is non-relativistic and either stable or has a very long lifetime, with a known number density $n_2$. 
Since each $X_2$ could decay into $X_1$ and a photon, the number density of photons $n_\gamma$ is proportional to the number density of the decaying particle $X_2$. Therefore, the expected energy density of photons is given by  
\begin{equation}
\rho_\gamma = E_\gamma n_\gamma \propto \frac{\Delta m^2}{2 m_2} \, n_2 \, .
\end{equation} 
The equality is governed by how many $X_2$ particles decay within a given time interval $\mathcal{T}$. 
Assuming that the $X_2$ lifetime is long enough, we find that $\rho_\gamma$ is given by
\begin{equation}
    \rho_\gamma = \Gamma \,\mathcal{T} n_2 \frac{\Delta m^2}{2 m_2}  \, ,
\end{equation}
where $\Gamma$ denotes the decay rate of $X_2$. Since our goal is to detect such decay photon signals using quantum devices, the time interval $\mathcal{T}$ should be less than the coherence time $\tau$ of a given quantum system, i.e., $\mathcal{T} < \tau$. 

As described earlier, we assume that the produced photon is well described by Eq.~\eqref{eq:efffield}. Given that the photon energy density stored in the electric field is $\rho_\gamma = \varepsilon_0 \vec{E}^2$, we obtain 
\begin{equation}
    \bar E_{\rm (eff)} = \sqrt{\frac{\Gamma \, \mathcal{T} n_2 \Delta m^2}{2  m_2 \varepsilon_0}} \, , \label{eq;Eeff}
\end{equation}
where $\varepsilon_0$ denotes the vacuum permittivity.\footnote{In natural units adopted here, $\varepsilon_0$ is dimensionless and set to unity.}
We emphasize that the effective electric field introduced above is a phenomenological representation of the energy deposited by an individual decay photon, rather than the field of a classical electromagnetic wave. Throughout this work, the term single photon refers to the energy transferred in a single decay event. For the purpose of estimating the qubit response, this energy transfer is represented by an effective monochromatic electric field acting on the quantum sensor.

This treatment is consistent with the fact that the qubit response is determined by the field two-point correlation function, rather than by the expectation value of the (electric) field operator $\hat{E}$, even when the photon occupation number in the cavity is small \cite{Glauber1963a,Glauber1963b}. In a cavity, these field correlation functions inherit the oscillatory time dependence set by the cavity eigenfrequencies.
For example, in a single-mode model, one finds that two-point functions contain harmonic factors at $\pm E_\gamma$ even for few-photon Fock states \cite{ScullyZubairy}.
In our setup, the relevant detector response is the absorption channel of a two-level system; within the rotating-wave approximation, this reduces the second-order response to the normally ordered first-order coherence function
$G^{(1)}(t_1,t_2)=\langle \hat{E}^{(+)}(t_1)\hat{E}^{(-)}(t_2)\rangle$.
Consequently, to leading order in perturbation theory, the excitation probability is governed by the near-resonant spectral weight of the field at the qubit transition frequency, consistent with the standard Unruh-DeWitt detector response formalism~\cite{Unruh1976,BirrellDavies}. A classical monochromatic field $\vec{E}(t)=\vec{E} \sin(E_\gamma t)$ possesses similar spectral support, and therefore produces the same transition probability up to normalization by the total mode energy \cite{Glauber1963a,ScullyZubairy}. For this reason, modeling the decay photon quantum as a sinusoidal field with energy $E_\gamma$ correctly captures the detector response even in the few-photon limit. A more detailed discussion can be found in Section~\ref{sec:appA}.

\subsection{Interaction with quantum devices}
\label{sec:devices}

We consider two types of quantum devices: ($a$) transmon qubits and ($b$) trapped ions. Brief overviews of these platforms follow here based on Refs. \cite{Chen:2022quj,Shu:2024nmc}.
\medskip

\noindent \textbf{($a$) Transmon qubits:} The effective electric field induced by the photon from the $X_2$ decay modifies the Hamiltonian of the transmon qubit by adding the following interaction term
\begin{eqnarray}
    {\mathcal H}_{\rm int} &=& C V d  \bar E_{\rm (eff)} \cos\Theta \sin (E_\gamma t) \\
    &=& 2 \eta \sin (E_\gamma t) \left(\hat a + \hat a^\dagger \right) \, ,  \label{eq:Hinoperator}   
\end{eqnarray}
where $C$ is the capacitance, $V$ is the voltage across the Josephson junction, $d$ is the effective distance between the two conductor plates, and the angle $\Theta$ is the angle between $\vec n_E$ and the normal vector of the conductor plate \cite{Chen:2022quj}. 
Again, the interaction above should be viewed as a response-level effective Hamiltonian obtained after matching the quantum-optics absorption probability to the near-resonant two-level response.  It is not an assumption that the decay photons form a coherent classical drive. 
The $\hat a$ and $\hat a^\dagger$ are the ladder operators associated with the ground and excited states $|g\rangle$ and $|e\rangle$ of the transmon, when approximating the non-linear potential by a Josephson junction in the transmon qubit system as a parabolic potential \cite{Chen:2022quj}. In this limit, the initial Hamiltonian of the system is reduced to approximately ${\mathcal H}_0 = \omega |e\rangle \langle e |$ with the energy of the ground state $|g\rangle$ taken to be zero. Here $\omega$ is the energy difference between the ground state $|g\rangle$ and the first excited state $|e\rangle$ of the quantum system. 
The coefficient $\eta$ for the transmon qubits is defined as
\begin{equation}
    \eta \equiv \frac{1}{2\sqrt 2} d \sqrt{C\omega} \cos\Theta \bar E_{\rm (eff)} \, .
\end{equation}
The resulting total Hamiltonian $\mathcal{H}=\mathcal{H}_0+\mathcal{H}_{\rm int}$ is given by
\begin{equation}
    {\mathcal H} = \omega |e\rangle\langle e | + 2 \, \eta \, \sin ( E_\gamma t) \, \big (
|e\rangle\langle g | + |g\rangle\langle e |
    \big ).
\end{equation}
This expression provides the phenomenological description of the interaction between the transmon qubits and the induced effective electric field, which leads to 
the transition probability from the ground state to the excited state
\begin{equation}
p_{g\to e}(t) = |\psi_e (t)|^2 \simeq \sin^2 \eta t \, , \label{eq:pstar_full}
\end{equation}
within the coherence time, $t < \tau$.  Assuming $\tau \ll \eta^{-1}$, the transition probability within the coherence time can be computed as 
\begin{equation}
p_\ast \equiv p_{g \to e}(\tau) \simeq (\eta\tau)^2 \, . \label{eq:transprob}
\end{equation}

We remark that although Eq.~\eqref{eq:pstar_full} has the familiar Rabi form, the sensitivity estimates in this work rely only on the weak-response limit in Eq.~\eqref{eq:transprob}. This limit is equivalent to the leading-order absorption probability derived from the normally ordered field correlator in Section~\ref{sec:appA} and is therefore applicable to an incoherent narrow-band decay-photon population.

\medskip
\noindent \textbf{($b$) Trapped ions:} The interaction between trapped ions and electromagnetic fields provides a fundamentally different sensing mechanism compared to transmon qubits. In trapped ion systems, the electromagnetic field from particle decays couples directly to the ion's mechanical motion through the Coulomb interaction. The position-dependent coupling arises because the ion's charge distribution oscillates in the harmonic potential created by the trap, making it sensitive to external electric fields. The free Hamiltonian of the trapped ion acts as a harmonic oscillator. The interaction Hamiltonian $\mathcal{H}_{\rm int}$ is given by
\begin{eqnarray}
    {\mathcal H}_{\rm int} &=& e \vec E_{\rm (eff)} \cdot \vec x \\
    &=&  \sum_n \frac{e E^n_{\rm (eff)}}{\sqrt{2 m_{\rm ion} w_r^n}} \hat a_n^\dagger e^{-i \omega_r^n t} + h.c., 
\end{eqnarray}
where $\vec x$ is the three-dimensional position operator, $E_{\rm (eff)}$ is the induced electric field at the location of the ion, $ n$ denotes the $n$th vibrational mode of the ion and $m_{\rm ion}$ is the ion mass \cite{Shu:2024nmc}. The $\eta$ parameter in the transition probability expressed in Eq.~\eqref{eq:transprob} 
is replaced with  
\begin{equation}
    \eta \equiv \frac{1}{2\sqrt 2} \frac{e \bar E_{\rm (eff)}}{\sqrt{m_{\rm ion} w_r^n}} \,  ,
\end{equation}
for trapped ions. The decaying particle couples resonantly to the sensor when $\omega_r^n = \omega^\ast = E_\gamma $, where $\omega^\ast =2\pi f^\ast$ is the characteristic frequency of the trapped ion device and $E_\gamma$ is the energy of the induced photon in the neutrino or DM decay. The specific allowed frequency range and coherence time depend on the type of ion used in the experiment. Typically, trapped ion systems have a significantly longer coherence time compared to superconducting qubits, allowing longer integration times and potentially higher sensitivity for weak and slowly varying signals. On the other hand, the coupling parameter is typically smaller due to the large ion mass $m_{\rm ion}$.

\section{Sinusoidal approximation from the two-point function}
\label{sec:appA}

We describe how a decay-photon quantum can be modeled, for the purpose of estimating the qubit excitation probability, as an effective near-monochromatic (sinusoidal) drive.
For a single-photon Fock state, and more generally for incoherent number-state-like photon populations relevant to randomly emitted decay photons, the one-point expectation value of the electric-field operator $\hat{E}(t)$ vanishes. For example, for a single-photon Fock state $|1\rangle$,
\begin{equation}
\langle \hat{E}(t)\rangle =\langle 1 | \hat{E}(t) | 1 \rangle = 0,
\end{equation}
so the field carries no classical amplitude in the one-point sense.
Nevertheless, the detector response is governed by field two-point correlations~\cite{Glauber1963a,Glauber1963b}. In particular, within the rotating-wave approximation, the relevant correlator for absorption-induced excitation is the normally ordered first-order coherence function (Glauber correlator),
$G^{(1)}(t_1,t_2)\equiv
\left\langle \hat{E}^{(+)}(t_1)\,\hat{E}^{(-)}(t_2)\right\rangle$,
rather than $\langle \hat{E}(t_1)\hat{E}(t_2)\rangle$. The latter contains vacuum and counter-rotating contributions that do not govern resonant absorption in the rotating-wave approximation.
Below, we present a general mode expansion for $G^{(1)}$ and show how it motivates a sinusoidal effective field near resonance.

\subsection{Qubit-field interaction and detector response}

We consider a linear (dipole-like) coupling between the qubit and the electric field at the qubit location,
\begin{eqnarray}
H_{\rm int}(t)
&=& g\,\hat{\sigma}_x(t)\,\hat{E}(t)
\nonumber\\
&=& g\left(\hat{\sigma}_- e^{-i\omega t}+\hat{\sigma}_+ e^{+i\omega t}\right)\hat{E}(t),
\label{eq:Hint_sigmaE}
\end{eqnarray}
where $g$ is a coupling constant and $\omega$ is the qubit transition frequency. Here $\hat{\sigma}_\pm =\frac{1}{2}(\hat{\sigma}_x\pm i\hat{\sigma}_y)$ generate $|g\rangle\to |e\rangle$ $(+)$ and $|e\rangle\to |g\rangle$ $(-)$ transitions. Therefore, the ladder operators $\hat{a}^\dagger$ and $\hat{a}$ in Eq.~\eqref{eq:Hinoperator} may be identified as $\hat{\sigma}_+$ and $\hat{\sigma}_-$, respectively.
In the rotating-wave approximation, we keep only the energy-conserving absorption term,
\begin{equation}
H_{\rm int}^{\rm (RWA)}(t)\simeq g\,\hat{\sigma}_+(t)\,\hat{E}^{(-)}(t)+{\rm h.c.},
\label{eq:Hint_RWA}
\end{equation}
where $\hat{E}^{(+)}$ ($\hat{E}^{(-)}$) denotes the positive- (negative-) frequency part of the field operator.

To leading order in time-dependent perturbation theory, the excitation amplitude from $|g\rangle$ to $|e\rangle$ is
\begin{eqnarray}
\mathcal{A}
&=&-ig\int_{0}^{T}dt\,
\langle e,f| \hat{\sigma}_+(t)\,\hat{E}^{(-)}(t)|g,i\rangle
\nonumber\\
&=&-ig\int_{0}^{T}dt\,e^{+i\omega t}\,\langle f|\hat{E}^{(-)}(t)|i\rangle,
\end{eqnarray}
where $|i\rangle$ and $|f\rangle$ denote the initial and final photon states.
Summing over unobserved final photon states and using completeness yields the excitation probability
\begin{widetext}
\begin{eqnarray}
p_{g\to e}(T)
=\sum_f|\mathcal{A}|^2 
&=& 
g^2 \int_0^T dt_1 \int_0^T dt_2 \,
e^{-i\omega (t_1 - t_2)}
\,
\Big\langle
\hat{E}^{(+)}(t_1)\hat{E}^{(-)}(t_2)
\Big\rangle
\nonumber\\
&\equiv& 
g^2 \int_0^T dt_1 \int_0^T dt_2 \,
e^{-i\omega (t_1 - t_2)}\,G^{(1)}(t_1,t_2),
\label{eq:Pg_e_general_G1}
\end{eqnarray}
\end{widetext}
which is the standard quantum-optics expression for absorption-driven excitation~\cite{Glauber1963a,Glauber1963b,ScullyZubairy}. In other words, starting from the full $\hat{\sigma}_x \hat{E}$ coupling, the second-order detector response can be written in terms of $\langle \hat{E}(t_1)\hat{E}(t_2)\rangle$; however, for excitation via absorption and within the rotating-wave approximation, the relevant contribution reduces to the normally ordered correlator $G^{(1)}(t_1,t_2)=\langle \hat{E}^{(+)}(t_1)\hat{E}^{(-)}(t_2)\rangle$ of quantum optics. In what follows, we exploit this correlator-level description to motivate an effective 
detector-response-level parameterization for estimating $p_{g\to e}$, without claiming equivalence to a classical field at the level of the full photon statistics.

\subsection{Single-photon field in a reflective cavity: general mode expansion}

Inside a reflective cavity, the electric field at the qubit location can be expanded in discrete cavity modes. Writing the negative- and positive-frequency parts,
\begin{eqnarray}
\hat{E}^{(-)}(t)
&\equiv&\sum_k \mathcal{E}_k\,\hat{a}_k\,e^{-i\omega_k t}, \nonumber \\
\hat{E}^{(+)}(t)&\equiv&\left(\hat{E}^{(-)}(t)\right)^\dagger
=\sum_k \mathcal{E}_k^*\,\hat{a}_k^\dagger\,e^{+i\omega_k t},
\label{eq:Eplus_multimode}
\end{eqnarray}
where $\omega_k$ are the cavity eigenfrequencies and $\hat{a}_k$ are annihilation operators for the corresponding modes. The coefficients $\mathcal{E}_k$ encode the mode function evaluated at the qubit position, polarization projection, and the zero-point amplitude (schematically $\mathcal{E}_k\sim \sqrt{\omega_k}\,u_k(\mathbf{r}_q)$ with $u_k$ denoting the normalized mode profile at the qubit location).
We emphasize that Eq.~\eqref{eq:Eplus_multimode} is general and does not assume a single-mode description.

Using Eq.~\eqref{eq:Eplus_multimode}, the first-order correlator becomes
\begin{equation}
G^{(1)}(t_1,t_2)
=
\sum_{k,k'}
\mathcal{E}_k^*\mathcal{E}_{k'}
\,e^{+i\omega_k t_1}e^{-i\omega_{k'} t_2}\,
\left\langle \hat{a}_k^\dagger \hat{a}_{k'}\right\rangle.
\label{eq:G1_multimode_general}
\end{equation}
Two limiting cases are useful:

\paragraph{Mode-diagonal number states.}
If the field state is a product of Fock states, $\rho=\bigotimes_k |n_k\rangle\langle n_k|$, then $\langle \hat{a}_k^\dagger \hat{a}_{k'}\rangle=n_k\,\delta_{kk'}$ and
\begin{equation}
G^{(1)}(t_1,t_2)
=
\sum_k |\mathcal{E}_k|^2\,n_k\,e^{-i\omega_k(t_2-t_1)}.
\label{eq:G1_multimode_diagonal}
\end{equation}
In particular, for a single mode $k=c$ populated with $n$ photons ($n_c=n$, $n_{k\neq c}=0$), one finds $G^{(1)}(t_1,t_2)=|\mathcal{E}_c|^2 n\,e^{-i\omega_c(t_2-t_1)}$. 
Note that only $n$ (not $n+1$) appears in the absorption correlator, as expected for an $n$-photon state. This limiting case illustrates that a nonzero absorption response does not require a coherent classical field amplitude; for an incoherent photon population, the detector response is instead controlled by the occupation-weighted local field intensity appearing in \(G^{(1)}\). 

\paragraph{Single-photon wavepacket across cavity modes.}
A decay photon generated inside the cavity is generically described by a superposition
over cavity modes,
\begin{equation}
|1_\alpha\rangle \equiv \sum_k \alpha_k\,\hat{a}_k^\dagger|0\rangle,
\qquad
\sum_k|\alpha_k|^2=1,
\label{eq:one_photon_wavepacket}
\end{equation}
for which
$\langle \hat{a}_k^\dagger \hat{a}_{k'}\rangle=\alpha_k^*\alpha_{k'}$, and thus
\begin{equation}
G^{(1)}(t_1,t_2)
=
\left[\sum_k \mathcal{E}_k^*\alpha_k^* e^{+i\omega_k t_1}\right]
\left[\sum_{k'} \mathcal{E}_{k'}\alpha_{k'} e^{-i\omega_{k'} t_2}\right].
\label{eq:G1_wavepacket_factorized}
\end{equation}
Equation~\eqref{eq:G1_wavepacket_factorized} shows explicitly that, in general, the correlator need not be a function of $(t_1-t_2)$ alone; rather, it is set by the temporal structure (wavepacket) of the single-photon state.

\subsection{Near-monochromatic limit and effective sinusoidal drive}

The qubit excitation probability in Eq.~\eqref{eq:Pg_e_general_G1} is controlled by the spectral overlap between the qubit transition and the field correlator. If the photon state has dominant support around a characteristic energy $E_\gamma$ and the cavity filters the field so that the effective local amplitude is narrowband, then the mode sum in Eq.~\eqref{eq:G1_wavepacket_factorized} can be approximated as
\begin{equation}
\sum_k \mathcal{E}_k \alpha_k e^{-i\omega_k t}
\;\simeq\;
\mathcal{E}_{\rm eff}\,f(t)\,e^{-iE_\gamma t},
\label{eq:narrowband_ansatz}
\end{equation}
where $f(t)$ is a slowly varying envelope (set by the photon wavepacket and/or the cavity lifetime) and $\mathcal{E}_{\rm eff}$ is an effective root-mean-square field amplitude at the qubit location. In this narrow-band limit,
\begin{equation}
G^{(1)}(t_1,t_2)
\simeq
|\mathcal{E}_{\rm eff}|^2\, f^*(t_1)f(t_2)\,e^{-iE_\gamma(t_2-t_1)}.
\label{eq:G1_narrowband}
\end{equation}
Substituting Eq.~\eqref{eq:G1_narrowband} into Eq.~\eqref{eq:Pg_e_general_G1} yields
\begin{equation}
p_{g\to e}(T)
\simeq
g^2|\mathcal{E}_{\rm eff}|^2
\left|
\int_0^T dt\, f(t)\,e^{+i(\omega-E_\gamma)t}
\right|^2.
\label{eq:Pge_narrowband_general}
\end{equation}
For a sufficiently long coherence time compared to the optical period, the dominant response is resonant at $E_\gamma\simeq\omega$, and Eq.~\eqref{eq:Pge_narrowband_general} reduces to the familiar $\dfrac{\sin^2x}{x^2}$ line shape. In the idealized case of a quasi-stationary photon amplitude within a time window $\tau$ [i.e.,\ $f(t)\simeq 1$ for $0<t<\tau$], one finds
\begin{equation}
p_{g\to e}(\tau)
\simeq
g^2|\mathcal{E}_{\rm eff}|^2\,
\frac{\sin^2\!\left(\frac{(\omega-E_\gamma)\tau}{2}\right)}
{\left(\frac{\omega-E_\gamma}{2}\right)^2}
\;\xrightarrow[\ E_\gamma \to \omega \ ]{}\;
g^2|\mathcal{E}_{\rm eff}|^2\,\tau^2.
\label{eq:Pge_resonant_tau}
\end{equation}
Given that Eq.~\eqref{eq:transprob} leads to a probability of a similar form, this motivates a phenomenological description in which the single-photon state is modeled, at the level of detector response.
In other words, Eq.~\eqref{eq:Pge_resonant_tau} shows that, in the narrow-band limit, the excitation probability is controlled by the same resonant line shape as that obtained from a sinusoidal drive. This observation motivates the response-level parameterization 
\begin{equation}
\vec{E}_{\rm resp}(t) = \vec{E}_{\rm (eff)}\,\sin(E_\gamma t),
\label{eq:Ecl_eff}
\end{equation}
where \(E_{\rm eff}\) is fixed by the near-resonant local spectral weight of \(G^{(1)}\). We emphasize that \(\vec E_{\rm resp}(t)\) is not the expectation value \(\langle \hat{E}(t)\rangle\), nor does it imply that the decay photons form a coherent classical electromagnetic wave. For randomly emitted decay photons, different events generally carry no common phase, and their contributions add at the level of intensity or correlation functions rather than coherent amplitudes. 

Finally, comparing Eq.~\eqref{eq:Pge_resonant_tau} with Eq.~\eqref{eq:transprob}, we relate the effective coupling to transmon parameters as
\begin{equation}
g=\frac{1}{4}d\sqrt{C\omega}\cos\Theta,
\qquad
\mathcal{E}_{\rm eff}\equiv \bar{E}_{\rm (eff)}.
\end{equation}
We here define $\mathcal{E}_{\rm eff}$ operationally by matching the local spectral weight of $G^{(1)}$ near resonance; hence it agrees with $\bar{E}_{\rm (eff)}$ up to convention-dependent $\mathcal{O}(1)$ factors (e.g.,\ from mode normalization and the precise definition of positive/negative-frequency components). With this convention, the effective Rabi parameter in the main text may be identified as $\eta \equiv g\,|\mathcal{E}_{\rm eff}|$.

In summary, although the one-point expectation value vanishes for few-photon states, the normally-ordered two-point function $G^{(1)}(t_1,t_2)$ exhibits the oscillatory structure that controls resonant absorption. When the single-photon wavepacket is narrow-band around $E_\gamma$ and the qubit probes it over a coherence time $\tau$, the transition probability can be estimated using an effective sinusoidal drive with amplitude calibrated to the local spectral weight. This mapping holds at the level of the transition probability, not of the full field statistics~\cite{Glauber1963a,Glauber1963b,ScullyZubairy}.

\section{Measurement Protocol and Sensitivity of quantum devices}
\label{sec:analysis}

The detection strategy relies on measuring small excitation probabilities $p_*$ through a systematic counting experiment using one or more quantum sensors (transmon qubits or trapped ions) operating at a common resonant frequency. The measurement protocol follows a cyclic approach designed to accumulate statistical evidence for particle decay signatures,  based on the measurement protocol described in Ref.~\cite{Chen:2022quj}:
\begin{itemize}
\item[(i)] All quantum sensors are prepared in their ground state at $t = 0$ and allowed to evolve freely under the influence of potential decay-induced electromagnetic fields until the coherence time limit $t = \tau$ is reached.
\item[(ii)] A quantum measurement is performed on all sensors to determine their final states. This readout process requires approximately 100 nanoseconds for transmon qubits or microseconds to milliseconds for trapped ions-durations that are negligible compared to the coherence times involved.
\item[(iii)] For a given resonant frequency $\omega$, the initialization-evolution-readout sequence is repeated $n_{\rm rep}$ times. When measuring individual qubits, and requiring at least one to transition from the ground to the excited state, the total number of measurements is $N_{\rm try} = n_q \times n_{\rm rep}$, where $n_q$ represents the number of quantum sensors operated simultaneously \cite{Chen:2022quj}. 
In contrast, if entanglement is used to boost sensitivity, the measurement is performed on an ancilla qubit. In this case, the total number of measurements reduces to $N_{\rm try} = n_{\rm rep}$ \cite{Chen:2023swh}.
Assuming the decay-induced transitions occur independently across sensors, the expected number of signal events is 
\begin{equation}
    N_{\rm sig} = p_* \times N_{\rm try}.
    \label{eq:Nsig}
\end{equation}
 \item[(iv)] The entire measurement campaign is repeated across different sensor frequencies to probe various particle mass ranges and decay channels, as different decay photon energies will resonate with sensors tuned to corresponding frequencies. 
We consider scanning the frequency range from 1~GHz to 10~GHz. The relevant spectral resolution is set by the finite sensor--field interaction time and is parameterized as \(\delta f \equiv \dfrac{f}{Q_{\rm int}}\), where we define an interaction-time quality factor \(Q_{\rm int}\equiv \dfrac{\omega\tau_{\rm int}}{2\pi}\).\footnote{Here \(Q_{\rm int}\) should be distinguished from the intrinsic cavity quality factor \(Q_{\rm cav}\).}
For definiteness, and assuming \(Q_{\rm int}\) is approximately constant over the scan band, we take \(Q_{\rm int}=10^6\), which yields approximately \(2.3\times 10^6\) resolvable frequency bins across the band. Running continuously for one year (\(3.15\times 10^7\)\,s) then corresponds to \(\sim 14\)~s of integration time per bin. 
In our setup, \(\tau_{\rm int}\) can be limited by both the photon residence time in the reflective enclosure and the intrinsic sensor coherence, \(\tau_{\rm int}=\min(\tau_{\rm cav},\tau_{\rm qubit})\) (see Appendix~\ref{sec:photonresidence}). In the benchmark treatment adopted here, we follow Ref.~\cite{Chen:2022quj} and infer \(\tau_{\rm int}\) from the chosen \(Q_{\rm int}\) via \(\tau_{\rm int}=\dfrac{2\pi Q_{\rm int}}{\omega}\), which effectively encodes the limiting interrogation time in a single parameter.

\end{itemize}

\begin{figure}[t]
\centering
\includegraphics[width=0.48\textwidth]{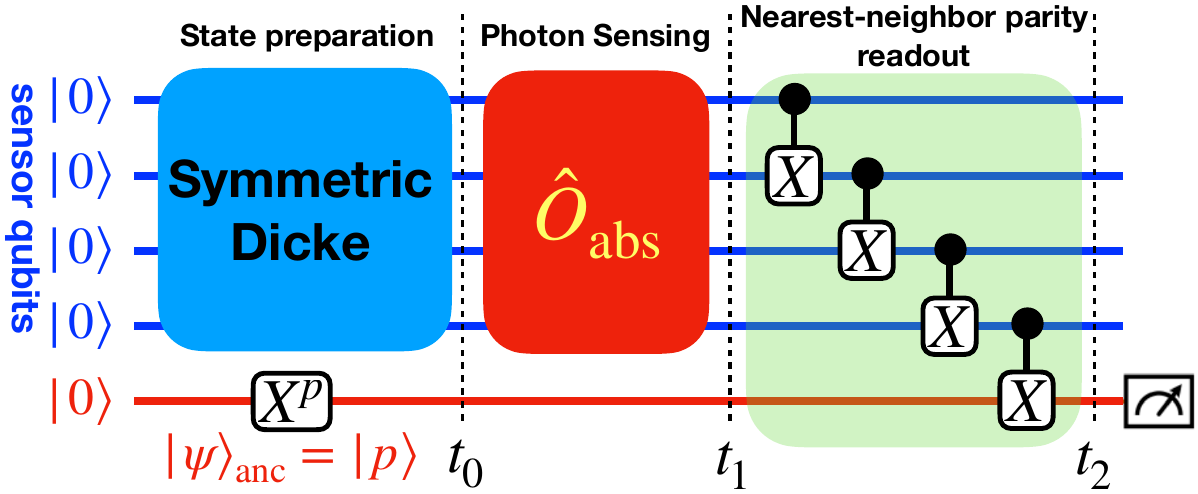}
\caption{Quantum circuit that can potentially yield an $\sim n_q^2$ collective enhancement in signal detection by preparing a symmetric Dicke state of $n_q$ qubit sensors. The total state factorizes into an ancilla register and the $n_q$-qubit sensor register: $|\Psi\rangle =|\psi\rangle_{\rm sensor} \otimes |\psi\rangle_{\rm anc}$. Here we prepare the ancilla qubit as a parity calibration with $p \equiv \dfrac{n_q}{2}~(\rm mod~2)$.  }
\label{fig:dicke} 
\end{figure}

This protocol effectively converts the quantum mechanical transition probability into a classical counting statistics problem, where the sensitivity depends on accumulating sufficient signal events above background to claim detection or set exclusion limits.

The false-positive excitations (also known as dark counts $N_{\rm dark}$) are critical in the sensitivity calculation. 
The primary source of dark counts is expected to be qubit readout error, given that state-of-the-art single-shot ground-state readout fidelity reaches approximately 99.9\% when utilizing higher excited states \cite{Chen:2022frn}. Accordingly, a uniform 0.1\% readout error is assumed in the subsequent discussion \cite{Chen:2022quj}. 
Thermal excitation is another important source of dark counts, which can be estimated as
\begin{eqnarray}
    N_{\rm dark} = e^{-\omega/T} N_{\rm try} \, ,
    \label{eq:Ndark}
\end{eqnarray}
where $T$ is the temperature of the transmon qubits or trapped ions. We will consider $T \sim 30~{\rm mK}$, when presenting our results. Other potential sources of false-positive excitations are shielded from the qubit, and are therefore considered negligible.

\section{Signal Enhancement via Dicke States}
\label{sec:dicke}
In the presence of multiple qubits, suitably engineered entanglement protocols can, in principle, yield a transition probability that scales as
\(p_\ast \simeq n_q^{2}(\eta\tau)^2\)~\cite{Chen:2023swh}.
The key point is that the readout circuit can preserve phase information such that signal amplitudes from different qubits add coherently, rather than contributing as independent, incoherent events. In this sense, \(n_q\) nominally weak probes can be combined into a single, effectively stronger measurement of the accumulated response. 

However, our target signal is a decay photon produced by rare DM decays, for which individual events are sporadic and do not naturally provide a stable phase reference across repeated trials. Consequently, it is unlikely that one can practically leverage any putative coherence associated with a wave-like DM field to enhance the decay-photon signal in an $n_q^2$-type manner.
Separately, for single-photon absorption, it is known that a collection of two-level systems prepared in appropriately engineered Dicke states \cite{Dicke:1954zz} can exhibit an enhanced absorption rate (also known as superabsorption) in idealized settings~\cite{Higgins2014}. 

While a detailed assessment of whether such Dicke-state engineering can be realized in the presence of amplitude-damping errors~\cite{He:2025ovo} is beyond the scope of this work, we briefly argue for the possible benefit of superabsorption-type enhancement and present a simple quantum circuit in Fig.~\ref{fig:dicke} that may realize this algorithm. 
More generally, Dicke-type collective states of $n_q$ qubits can be prepared using exchange-type two-qubit gates such as partial iSWAP operations arranged in a suitable circuit architecture~\cite{Saleem:2023vse}.
In particular, the half-excited symmetric Dicke state at a reference time $t_0$,
\begin{equation}
    |\psi(t_0)\rangle_{\rm sensor}=\left|J=\frac{n_q}{2}, m=0\right\rangle
\end{equation} 
can be generated given the ability to implement the required multi-qubit entangling sequence. 
Thus, the total state of this circuit becomes 
\begin{equation} 
|\Psi(t_0)\rangle = \left |\frac{n_q}{2}, 0\right\rangle\otimes 
|p\rangle,
\end{equation}
where $|p\rangle$ represents the ancilla qubit state with a parity calibration as $p \equiv \dfrac{n_q}{2}\, (\rm mod~2)$. 
For the $n_q$-qubit entangled system, the interaction Hamiltonian in Eq.~\eqref{eq:Hinoperator} becomes
\begin{equation}
    \mathcal{H}_{\rm int}^{(\{n_q\})}= 2\eta \sin(E_\gamma t)\sum_{i=1}^{n_q} \hat{\sigma}_x^{(i)}= 2\eta\sin(E_\gamma t)(\hat{J}_+ + \hat{J}_-),
\end{equation}
where we identify the two-state ladder operator $\hat{a}$ and $\hat{a}^\dagger$ in Eq.~\eqref{eq:Hinoperator} by $\hat{\sigma}_{-}$ and $\hat{\sigma}_{+}$, hence $\hat{a}+\hat{a}^\dagger\to \hat{\sigma}_x$. 
Now, the photon-absorption transition is given by
\begin{eqnarray}
    |\Psi(t_1)\rangle &\approx& \hat{\mathcal{O}
    }_{\rm abs} |\Psi(t_0)\rangle \\
    &=& \left( \left|\frac{n_q}{2}, 0\right\rangle- i (\eta \, t_1) \sqrt{\frac{n_q}{2}\left(\frac{n_q}{2}+1\right)} \left|\frac{n_q}{2}, 1\right\rangle \right)\otimes |p\rangle, \nonumber 
\end{eqnarray}
where $\hat{\mathcal{O}}_{\rm abs}\approx I-i(\eta t)\hat{J}_+$ at the leading order with $t_0 < t_1 \lesssim t_2 \lesssim \tau$. A  chain of nearest-neighbor CNOT-gate operations $U_{\rm CX}=\left(\prod_{i=1}^{n_q-1} {\rm CX}_{i, i+1}\right) {\rm CX}_{n_q, n_q+1}$ now yields
\begin{eqnarray}
    &&|\Psi(t_2)\rangle =\left(\prod_{i=1}^{n_q-1} {\rm CX}_{i, i+1}\right) {\rm CX}_{n_q, n_q+1} |\Psi(t_1)\rangle \nonumber \\
    &&= \left(\prod_{i=1}^{n_q-1} {\rm CX}_{i, i+1}\right)\left|\frac{n_q}{2},0\right\rangle \otimes |0\rangle \\ 
    &&- i(\eta \, t_1) \sqrt{\frac{n_q}{2}\left(\frac{n_q}{2}+1\right)}\left(\prod_{i=1}^{n_q-1} {\rm CX}_{i, i+1}\right)\left|\frac{n_q}{2}, 1\right\rangle 
\otimes |1\rangle, \nonumber
\end{eqnarray}
Thus, the probability that the ancilla is measured in the flipped state $|1\rangle$ is proportional to $\dfrac{n_q^2}{4}+\dfrac{n_q}{2}$ (i.e., $n_q^2$ scaling behavior in the large $n_q$ limit), due to the collective (Dicke) enhancement of the many-body transition matrix element for single-photon absorption within the symmetric Dicke states.
However, implementing an idealized Dicke-state superabsorption protocol in practice can be challenging on near-term quantum hardware due to intrinsic noise sources, with amplitude damping expected to be a particularly relevant limitation~\cite{He:2025ovo}.

\section{Sensitivity of quantum devices to physics models}
\label{sec:sensitivity}

We now illustrate the search potential of quantum devices for physics signals, beginning with the radiatively decaying DM scenario, followed by the C$\nu$B decay.

\subsection{Dark matter decay}
\label{sec:dm}

In this paper, we consider a two-component DM system $X_i$ ($i=1, 2$) with $X_1$ and $X_2$ representing the lighter and heavier DM candidates of masses $m_1$ and $m_2$, respectively.  Several spin assignments for $X_i$ may be considered. In the case of (pseudo)scalar $X_i$, example interactions include a dimension-6 operator, $\partial_\mu X_1 \partial_\nu X_2 F^{\mu\nu}$ with $F^{\mu\nu}$ being the usual field strength tensor for the SM photon. However, angular momentum conservation forbids a scalar $X_2$ from decaying into $X_1$ and a photon. If a different spin is assigned to $X_1$, such a decay is allowed---for example, through a dark (pseudo)scalar portal-type operator of the form $X_2 F_{\mu\nu}X_1^{\mu\nu}$~\cite{deNiverville:2018hrc}, where $X_1^{\mu\nu}$ is the field strength of $X_1$.  
If $X_i$ is a fermion, $m_2$ is required to be greater than $\sim 100$~eV in order to remain consistent with the Pauli exclusion principle~\cite{Tremaine:1979we}. 
In this case, quantum devices with ${\mathcal O}(1) {\rm GHz}$ operating frequency are insensitive to the associated photon signal.
However, this mass bound may be avoided by introducing multiple distinct DM components with quasi-degenerate masses and no couplings to the SM particles~\cite{Davoudiasl:2020uig}. 

In this subsection, for illustration, let us consider the following interaction between massive spin-1 $X_i$ and the photon, 
\begin{equation}
   \mu X_1^\mu X_2^\nu \tilde{F}_{\mu\nu},
\end{equation}
where $\mu$ denotes the dimensionless coupling strength. 
The full gauge-invariant set of interactions conceiving this operator can be found in Refs.~\cite{Lee:2015zqz,Dudas:2009uq,Anastasopoulos:2006cz}.\footnote{While we take this particular interaction as an illustration, one can consider any model where the heavier DM decays to the lighter one and a photon.} The decay rate of the heavier DM $X_2$ into the lighter one $X_1$ and a photon is computed by 
\begin{eqnarray}
    \Gamma(X_2\to X_1+\gamma)&=&\frac{\mu^2}{96 \pi}\frac{(m_2^2-m_1^2)^3(m_2^2+m_1^2)}{m_2^5m_1^2} ~~\\
    &\approx& \frac{\mu^2}{48\pi}\frac{(\Delta m^2)^3}{m_2^5}, \label{eq:nu_decay}
\end{eqnarray}
where the approximation in the second line is valid for $\Delta m^2 \ll m_1^2 \lesssim m_2^2  $.

While the relative abundances of $X_1$ and $X_2$ are model-dependent, for illustration, we take $X_2$ as the dominant component and $X_1$ as subdominant 
\begin{equation}
    \rho_{\rm DM} \approx m_2 \, n_{2}\,, \label{eq:rhoDM}
\end{equation} 
where $n_2$ is the number density of $X_2$. 
To ensure that $X_2$ has a sufficiently long lifetime, we impose the condition $\Gamma \lesssim \Gamma_{\rm U} = 2.299 \times 10^{-18} \, {\rm s^{-1}}$, which yields the following upper bound on the coupling $\mu$ (in the limit of $\Delta m^2 \ll m_1^2 \lesssim m_2^2  $)
\begin{eqnarray}
\big (  \mu_{\rm \max}^{\rm Uni} \big )^2 \approx 48 \pi \, m_2^2 \, \Big ( \frac{m_2}{\Delta m^2}\Big )^3 \, \Gamma_{\rm U}  \, .
\label{eq:mumax} 
\end{eqnarray}
Under this condition (i.e., $\Gamma\to \Gamma_{\rm U}$ or equivalently, $\mu \to\mu_{\max}^{\rm Uni}$), the transition probability for the transmon qubit device becomes
\begin{widetext}
\begin{align}
  p_\ast  &\simeq \, p_\ast^0  \times \cos^2 \Theta
  \left ( \frac{ 
  \rho_{\rm DM}}{0.45 \,{\rm GeV \, cm^{-3}}} \right )
  \left ( \frac{\Delta m^2}{m_2^2}\right)
\left( \frac{d}{100\ \mu{\rm m}} \right)^2
  \left( \frac{C}{0.1\ {\rm pF}} \right)  
  \left( \frac{f}{1\ {\rm GHz}} \right)
  \left( \frac{\tau}{100\ \mu{\rm s}} \right)^3 \, ,
\label{eq:p_ge_DM1}
\end{align}
\end{widetext}
where $p_\ast^0$ denotes a reference transition probability, evaluated to be $p_\ast^0 = 0.069$. 
Note that the characteristic frequency determined by the photon energy is not an independent parameter but is related to the DM masses via Eq.~\eqref{eq:Egamma}: 
\begin{eqnarray}
  f \simeq 1.2 \ {\rm GHz} \times
  \left( \frac{\Delta m^2} {10^{-10}\ {\rm eV^2}} \right) \left ( \frac{\rm 10^{-5}\ eV}{m_2} \right ) \, .
  \label{eq:DMfrequency}
\end{eqnarray} 
Constraints on DM decaying into photons at radio frequencies [$\mathcal{O}(1)$ GHz] are primarily derived from CMB spectral distortions \cite{Acharya:2023bln,Chluba:2015hma,Arsenadze:2024ywr} and the radio background observation \cite{Caputo:2022keo,Bernal:2020lkd}. Decays occurring at early times will primarily be constrained by distortions, while decays at late times to today will be heavily constrained by radio surveys and their potential to heat the intergalactic medium, which disrupts 21-cm signals. Computing such bounds is model-dependent and beyond the scope of this study. Instead, in this paper, we adopt a more stringent constraint on the DM lifetime based on CMB and reionization bounds, $\Gamma \lesssim \Gamma_{\rm CMB}=1.7 \times 10^{-25}~{\rm s^{-1}}$ \cite{Zhang:2007zzh}, which results in the significantly reduced reference probability $p_\ast^0 = 5.15 \times 10^{-9}$. 

\begin{figure*}[t]
    \centering
    \begin{minipage}[t]{0.48\textwidth}
        \centering
    \includegraphics[width=0.98\textwidth]{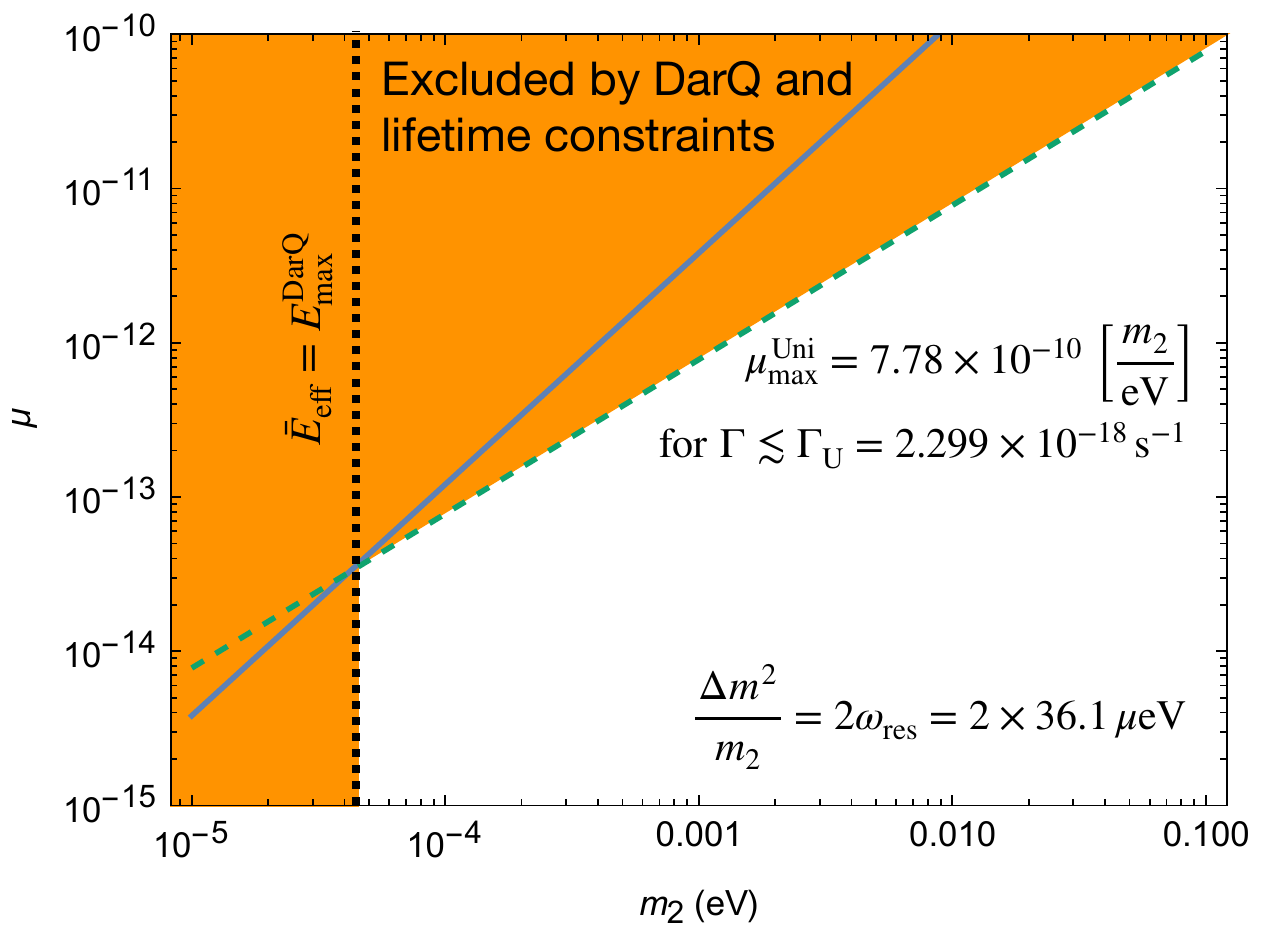}   
    \caption{Expected bounds on ($m_2$, $\mu$) by the DarQ experiment at 95\% CL \cite{Nakazono:2025tak}, which leads to $m_2 > 4.15\times 10^{-5} \, {\rm eV}$ for $\dfrac{\Delta m^2}{m_2} = 2 \omega_{\rm res}$ assuming $\Gamma \lesssim \Gamma_{\rm U} = 2.299 \times 10^{-18} \, {\rm s^{-1}}$ and coherence time of DarQ, $\mathcal{T}=1~\mu$s. Taking the CMB bound, $\Gamma \lesssim \Gamma_{\rm CMB}=1.7 \times 10^{-25}~{\rm s^{-1}}$, we obtain $m_2 > 3.07 \times 10^{-12} \, {\rm eV}$. 
    }
    \label{fig:DMbounds}
    \end{minipage}
    \hfill
    \begin{minipage}[t]{0.48\textwidth}
        \centering
    \includegraphics[width=0.99\textwidth]{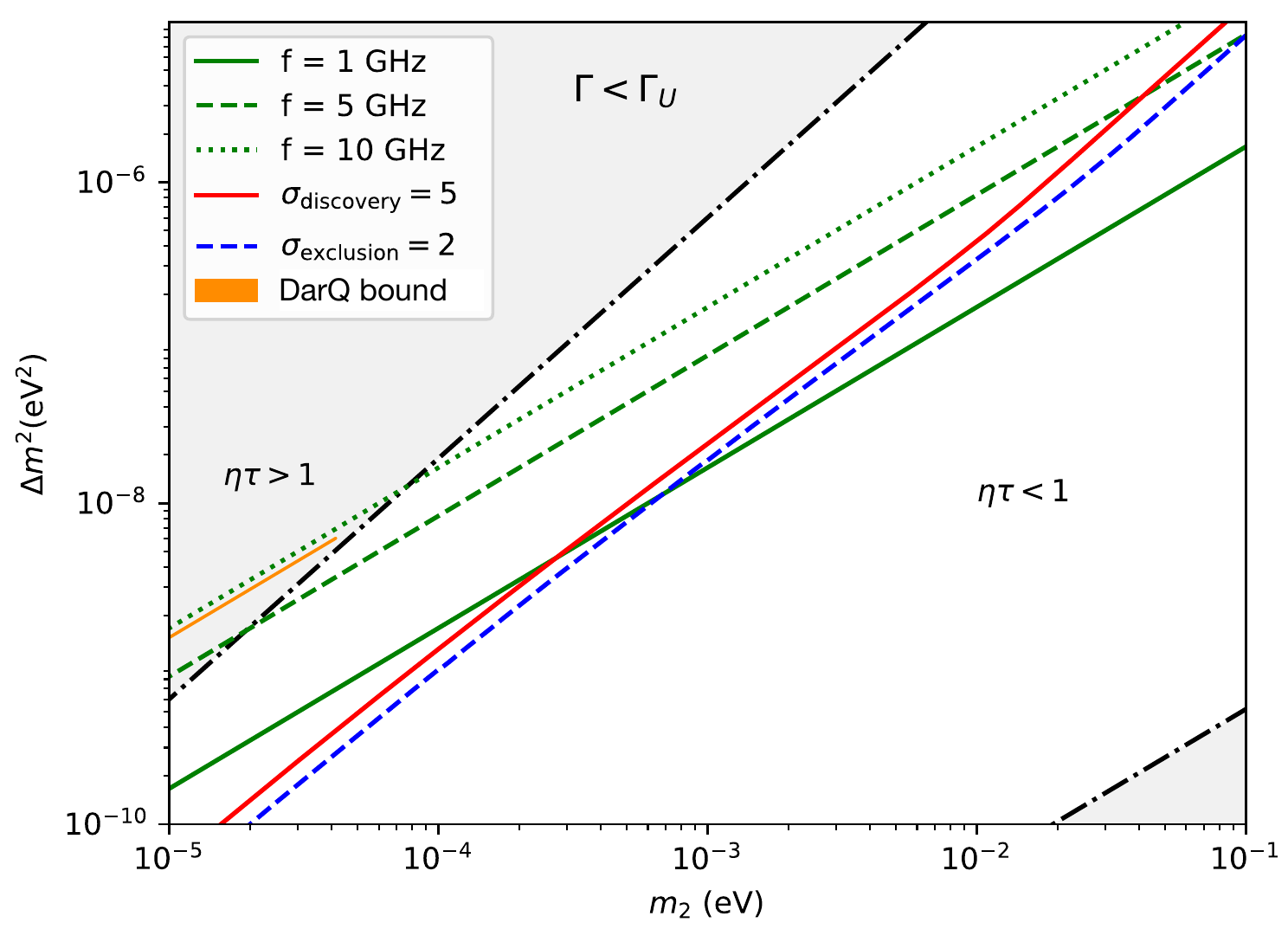}   
    \caption{The 5$\sigma$ discovery (red, solid) and 2$\sigma$ exclusion (blue, dashed) significance in the $(m_2, \Delta m^2)$ plane, assuming $\mu = \mu_{\max}^{\rm Uni}$, $\tau=100\,  \mu $s, $n_q=1$, and a 14-second data collection for each frequency. The orange-solid line represents the area excluded by the DarQ results. 
    The light-shaded region indicates the parameter space where $\eta\tau > 1$, requiring the use of the full expression $p_{g \to e} \approx \sin^2(\eta \tau)$.  
    \label{fig:DMbounds2} }
    \end{minipage}
\end{figure*}

The DarQ collaboration has recently reported its first results in the search for the hidden dark photon using transmon qubits~\cite{Nakazono:2025tak}. Although the related detection principle for hidden dark photons differs, the DarQ results can still constrain the parameter space of the decaying DM scenario under consideration, as they are effectively sensitive to $\bar{E}_{\rm (eff)}$.
DarQ shows that the kinetic mixing parameter $\epsilon$ must be below $10^{-12}$ for the resonance frequency $f_{\rm res}^{\rm DarQ}=\dfrac{\omega_{\rm res}^{\rm DarQ}}{2\pi}=$ 8.74 GHz. This resonance frequency corresponds to 36.1 $\mu \rm eV$, and by Eq.~\eqref{eq:Egamma} we identify
\begin{equation}
E_\gamma = \frac{\Delta m^2}{2 m_2} = \omega_{\rm res}^{\rm DarQ}= 36.1~\mu{\rm eV}. \label{eq:wresDarQ}
\end{equation}
Taking $\rho_{\rm DM}=0.45 \ \rm {GeV}/{cm^3}$, the expected limit on the effective electric field becomes $E_{\rm max}^{\rm DarQ} = \epsilon \sqrt{2 \rho_{\rm DM}} = 2.63 \times 10^{-15} \rm \ {eV^2}$. By requiring $\bar E_{(\rm eff)} < E_{\rm max}^{\rm DarQ}$, the exclusion limit based on the DarQ results \cite{Nakazono:2025tak} can be mapped onto the relevant parameter space for DM decay. 
Using Eqs.~\eqref{eq;Eeff}, \eqref{eq:rhoDM}, and \eqref{eq:wresDarQ}, 
we obtain the following bound: 
\begin{equation}
m_2 \geqslant
\frac{\Gamma \mathcal{T}\omega_{\rm res}^{\rm DarQ}}{2\varepsilon_0\epsilon^2}= 4.15\times 10^{-5}~{\rm eV},
\end{equation}
where we take 1~$\mu$s as the coherence time $\mathcal{T}$ of DarQ~\cite{TakeoMoroi} and assume $\Gamma=\Gamma_{\rm U}=2.299\times 10^{-18}~{\rm s}^{-1}$. This is represented by the vertical black dotted line in Fig.~\ref{fig:DMbounds}, to the left of which the corresponding $m_2$ values are excluded by DarQ. Under the same assumption, the upper bound on the $\mu$ parameter can be obtained using Eq.~\eqref{eq:mumax}, noting that $\dfrac{\Delta m^2}{m_2} = 2 \omega_{\rm res}^{\rm DarQ}$:
\begin{equation}
\mu \leqslant \mu_{\max}^{\rm Uni}= 7.78 \times 10^{-10}\times\left( \frac{m_2}{\rm eV} \right). 
\label{eq:mumaxDarQ} 
\end{equation}
This bound is shown as the dashed green line in Fig.~\ref{fig:DMbounds}, indicating that the region above this line is excluded by DarQ.
Finally, we present the constraint on $\mu$ in Eq.~\eqref{eq:nu_decay} as a function of the DM mass $m_2$, i.e., not fixing $\Gamma$. The allowed region is given by
\begin{equation}
    \mu^2 \leqslant \frac{12 \pi\epsilon^2\varepsilon_0}{(\omega_{\rm res}^{\rm DarQ})^4 \mathcal{T}}m_2^{3}=1.46\times 10^{-14}\times \left( \frac{m_2}{\rm eV}\right)^3\,.
\end{equation}
and the associated boundary is shown by the solid blue line.
In summary, the constraints from the DarQ results lead to a lower bound of $m_2 \geqslant 4.15 \times 10^{-5} \, {\rm eV}$ along the fixed-ratio slice $\dfrac{\Delta m^2}{m_2} = 2 \omega_{\rm res}^{\rm DarQ}$. Taking the CMB bound, $\Gamma \lesssim \Gamma_{\rm CMB}=1.7 \times 10^{-25}~{\rm s^{-1}}$ (instead of $\Gamma \lesssim \Gamma_{\rm U}$), we obtain $m_2 \geqslant  3.07 \times 10^{-12} \, {\rm eV}$ with $\mu_{\rm \max}^{\rm CMB}=2.12\times 10^{-13} \times \left(\dfrac{m_2}{\rm eV} \right)$.

Following the measurement protocol in Section \ref{sec:analysis}, in Fig. \ref{fig:DMbounds2}, we now show the 5$\sigma$ discovery (red, solid) and 2$\sigma$ exclusion\footnote{i.e., the line above is excluded, if there is no signal.} (blue, dashed) significances in the $(m_2, \Delta m^2)$ plane, assuming the maximum $\mu$ coupling in Eq.~\eqref{eq:mumax} and a 14-second data collection for each frequency as per the measurement protocol with a constant coherence time of 100 $\mu s$. For illustrative purposes, the scanned frequencies for the significance contours effectively cover the range starting from approximately 0.4~GHz and above. 
The $5\sigma$ discovery significance is calculated via the maximum likelihood ratio for excluding the background-only hypothesis~\cite{Cowan:2010js}:
\begin{eqnarray}
\sigma_{\rm discovery}=\sqrt{-2\ln\left(\frac{L(B|S+B)}{L(S+B|S+B)}\right)}\geq 5,
\label{eq:5sigma}
\end{eqnarray}
and the $2\sigma$ exclusion significance by excluding the signal plus background hypothesis:
\begin{eqnarray}
\sigma_{\rm exclusion}=\sqrt{-2\ln\left(\frac{L(S+B|B)}{L(B|B)}\right)}\geq 2,
\label{eq:2sigma}
\end{eqnarray}
where $S$ is the number of signal events [Eq.~\eqref{eq:Nsig}], $B$ is the number of background events, and $L$ is the Poisson distribution, 
$L(x|y)=\dfrac{x^y}{y!}e^{-x}$.
To compute the false-positive excitations, i.e., $B$, we use Eq.~\eqref{eq:Ndark} with a uniform 0.1\% readout error for the temperature, $T=30$~mK. We find that results for $T=1$~mK are similar to those for $T=30$~mK. 
While varying $m_2$ and $\Delta m^2$ over the parameter space---and thereby dynamically determining the characteristic frequency\footnote{whether or not the resulting frequency value is feasible for a given transmon qubit system.} by Eq.~\eqref{eq:DMfrequency}---to calculate these significances, we fix the remaining parameters as follows: 
$C=0.1\ {\rm pF}$, 
$d=100\ {\rm \mu m}$, 
$\rho_{\rm DM} = 0.45 \ {\rm GeV \ cm^{-3}}$, 
$\tau=100\ {\rm \mu s}$, and
$\cos^2\Theta=1/3$ as in Ref.~\cite{Chen:2022quj}.

As discussed in Section \ref{sec:analysis}, an experiment is conducted by fixing the characteristic frequency to a specific value. Unlike in the significance calculation, a fixed frequency probes a specific set of $(m_2,\Delta m^2)$ pairs, again as determined by Eq.~\eqref{eq:DMfrequency}.  
As representative examples, we display three frequency choices: 1 GHz (green, solid), 5 GHz (green, dashed), and 10 GHz (green, dotted). For example, if a transmon qubit system works between 1 GHz and 10 GHz, it essentially explores the band defined by the solid and dotted green lines. 
The DarQ exclusion bounds are shown in the (orange, solid) line. The light-shaded region marks the parameter space where $\eta\tau > 1$, in which case the transition probability must be evaluated using the full expression $p_{g \to e} \approx \sin^2(\eta \tau)$ as in Eq. (\ref{eq:pstar_full}). Two black dot-dashed lines represent the boundary defined by $\eta\tau=1$.

\bigskip

\begin{table*}[t]
\renewcommand\arraystretch{1.5}
\centering
\begin{tabular}{c||c|c|c|c|c|c}
\toprule
\hline \hline 
\textbf{~Ion Species~} & \textbf{~Transition Frequency ($f^\ast$)~} & ~$\tau^\ast$ (sec)~ & ~$m_2^\ast$ (eV)~ & ~$m_{\rm ion}$ (amu)~ & ~$p^0_\ast$ with $\Gamma_{\rm CMB}$~ & ~$p^0_\ast$ with $\Gamma_{\rm U}$~ \\
\hline \hline 
\midrule
$^{171}\mathrm{Yb}^+$ \cite{olmschenk2007manipulation,Wang_Luan_Qiao_Um_Zhang_Wang_Yuan_Gu_Zhang_Kim_2021} & \SI{12.6}{\giga\hertz} (2.38 cm)       &     10  & 1 & 170.9 & $9.8\times 10^{-11}$ & 0.0013 \\
$^{43}\mathrm{Ca}^+$ \cite{ballance2016high,harty2014high}    & \SI{3.2}{\giga\hertz} (9.4 cm)        &50 & 1 & 42.9588        & $4.9 \times 10^{-8}$ & 0.66 \\ 
\hline \hline 
\bottomrule
\end{tabular}
\caption{Trapped ion qubit parameters for two ion species in the case of DM decays. The starred quantities are the reference values adopted in our analysis.}
\label{table:trappedionDM}
\end{table*}

The transition probability $p_\ast$ for trapped ions in the case of DM decay can be calculated as 
\begin{align}
p_\ast &= 
\frac{1}{8} \left ( \frac{e \tau \bar E_{\rm (eff)}}{\sqrt{m_{\rm ion} \, \omega}} \right)^2 \\
&=  p_\ast^0 
\left ( \frac{ 
\rho_{\rm DM}}{0.45 \, \rm GeV \, cm^{-3}} \right )
\left ( \frac{\tau}{\tau^\ast} \right )^3
\left ( \frac{m_2^\ast}{m_2} \right )\,,
\end{align}
where the starred quantities indicate reference values. 
Table \ref{table:trappedionDM} summarizes the trapped ion qubit parameters for two ion species relevant to DM decays, including our reference value choices. 
We have chosen $m_2^\ast = 1$ eV for illustration purposes, but one can choose any values for the mass of the decaying DM. Once it is fixed, the mass-squared difference is given by $\Delta m^2 = 2 \omega^\ast m_2=4\pi f^\ast m_2$.
To determine $\tau^\ast$ used in computing $p^0_\ast$ in Table \ref{table:trappedionDM}, we adopt the shortest among the typical characteristic time scales, including the longitudinal relaxation time, transverse relaxation time, and inhomogeneous dephasing time.

\subsection{Cosmic neutrino background and transition magnetic moment}
\label{sec:cnb}

The detection of the C$\nu$B remains one of the most compelling challenges in modern physics, primarily due to its extremely low-energy nature~\cite{Muller:1987qm}. Despite significant theoretical and experimental efforts, including attempts via large-neutrino-mass scenarios \cite{Alvey:2021xmq,Perez-Gonzalez:2024xgb}, the PTOLEMY project \cite{PTOLEMY:2019hkd}, the dielectric response of boosted C$\nu$B-plasmon scattering \cite{Chao:2021ahl}, and quantum-induced broadening \cite{Nussinov:2021zrj}, the direct detection of the C$\nu$B has not been achieved yet.

As described previously, recent advancements in quantum technologies, particularly in quantum sensing \cite{Bass:2023hoi,Khan:2025ire}, present a promising new avenue for addressing this challenge. Superconducting transmon qubits, known for their sensitivity to weak interactions, have shown potential in the detection of DM within a mass range of $10^{-6}$ to $10^{-4}$ eV \cite{Chen:2022quj,Chen:2023swh,Chen:2024aya,Shu:2024nmc,Das:2022srn,Das:2024jdz}. Intriguingly, this energy range overlaps with the estimated energy spectrum of the C$\nu$B, which lies between $10^{-6}$ and $10^{-3}$ eV. Given the exceptionally weak interaction strength of DM with transmon qubits, it is plausible that similar techniques could be adapted for probing the C$\nu$B through its interactions with these devices. 

Furthermore, quantum entanglement and other quantum phenomena \cite{Ito:2023zhp} offer new opportunities for using quantum devices in particle physics experiments. While significant progress has been made in applying quantum technologies for DM detection, direct C$\nu$B detection with superconducting transmon qubits remains unexplored. In this section, we attempt to address this gap by examining the feasibility of detecting the C$\nu$B with quantum devices, after a brief review of the neutrino transition magnetic moment.

Current experimental values for the squared mass differences of three neutrinos are  
\begin{align}
\Delta m_{21}^2 &= 7.59 \times 10^{-5}\ {\rm eV^2} \, \\
\Delta m_{32}^2 &= 2.32 \times 10^{-3}\ {\rm eV^2} \, ,
\end{align}
which leads to two non-equivalent orderings for the neutrino masses. The normal hierarchical (NH) spectrum ($m_1 \ll m_2 < m_3$) and inverted hierarchical (IH) spectrum ($m_3 \ll m_1 < m_2$) lead to 
\begin{align}
m_2 &\simeq \sqrt{\Delta m_{21}^2} \sim 8.6 \times 10^{-3}\ {\rm eV} \, \\ 
m_3 &\simeq \sqrt{\Delta m_{32}^2 + \Delta m_{21}^2} \sim 0.05\ {\rm eV} \, ,
\end{align}
for NH and 
\begin{align}
m_1 &\simeq \sqrt{|\Delta m_{32}^2 + \Delta m_{21}^2|} \sim 0.0492\ {\rm eV} \, , \\
m_2 & \simeq \sqrt{|\Delta m_{32}^2|} \sim 0.05\ {\rm eV} \, ,
\end{align}
for IH, respectively \cite{ParticleDataGroup:2024cfk}. Therefore, the heavier C$\nu$B neutrinos are non-relativistic, considering $E_{\rm C\nu B} \sim 10^{-6} - 10^{-4}\ {\rm eV}$.

Electromagnetic properties of neutrinos have been extensively studied~\cite{Cisneros:1970nq,Kopp:2022cug,Fujikawa:1980yx,Kayser:1982br,Broggini:2012df,DOlivo:1989ued,Nieves:1981zt,Rosenberg:1962pp,Kouzakov:2017hbc,Giunti:2015gga,Bernstein:1963jp,MammenAbraham:2023psg}. In the simplest extension of the SM including the right-handed neutrinos, the neutrino magnetic moment is given by
\begin{eqnarray}
\mu_\nu = \frac{3 e G_F m_\nu}{8\sqrt{2}\pi^2} \approx 3 \times 10^{-19} \mu_B \left ( \frac{m_\nu}{1~{\rm eV}} \right ) \, 
\end{eqnarray}
where $m_\nu$ is the neutrino mass, $e$ is the electromagnetic coupling strength, $G_F$ is the Fermi constant, and $\mu_B = \dfrac{e}{2 m_e}$ is the Bohr magneton with $m_e$ being the electron mass \cite{Fujikawa:1980yx}. The neutrino magnetic moment interaction \cite{Giunti:2014ixa,Brdar:2020quo} is described by 
\begin{equation}
    \frac{\mu_\nu^{ij}}{2} \bar\nu_i \sigma^{\mu\nu} \nu_j \, F_{\mu\nu} + h.c.,
\end{equation}
where $i, j=1,2,3$. 

The SM prediction for the neutrino lifetime, based on the above interaction, exceeds the age of the Universe by many orders of magnitude, $\tau_{\rm SM} = 7.1 \times 10^{43} \left ( \dfrac{m_\nu}{\rm eV} \right )^{-5}$~seconds; see also Refs.~\cite{Hasegawa:2019jsa,Pal:1981rm,Petcov:1976ff}. However, various new physics scenarios beyond the SM can significantly enhance neutrino magnetic moments. Such modifications have been considered in connection with experimental anomalies, e.g., a possible correlation of solar neutrinos with solar activity~\cite{Bernal:2021ylz}. 
See also Refs.~\cite{Hasegawa:2019jsa,Pal:1981rm,Petcov:1976ff,Fujikawa:1980yx} for more details on the radiative decay of neutrinos and Refs.~\cite{Chupp:1989kx,Schonert:1995xh,Giunti:2014ixa} for experimental bounds.

The number density of cosmic neutrinos is given by 
\begin{equation}
    n_\nu = \frac{3}{4} \frac{\zeta(3)}{\pi^2} T_\nu^3 N_{\rm eff} \approx 112~{\rm cm^{-3}} \, ,
\end{equation}
for each flavor with $T_\nu=1.95 ~K$ and $N_{\rm eff}=3.046$. $\zeta(s) = \sum_{n=1}^\infty {1}/{n^s}$ is the Riemann zeta function. Following the discussion in Section \ref{sec:methods}, we obtain the effective electric field for the neutrino magnetic moment,
\begin{equation}
    \bar E_{\rm (eff)} = \sqrt{\frac{\Gamma_{ij} \mathcal{T} \, 
    n_\nu \Delta m_{ij}^2}{2 m_i \varepsilon_0}} \, , \label{eq:Eeff}
\end{equation}
where $\Gamma_{ij} = \Gamma (\nu_i \to \nu_j + \gamma)$, expressed as
\begin{equation}
\Gamma (\nu_i \to \nu_j + \gamma) = \frac{(\mu_\nu^{ij})^2}{8\pi m_i^3} \Big (m_i^2 - m_j^2 \Big )^3 \, . \label{eq:nudec}
\end{equation}

\begin{figure*}[th!]
\centering
\includegraphics[width=0.5\textwidth]{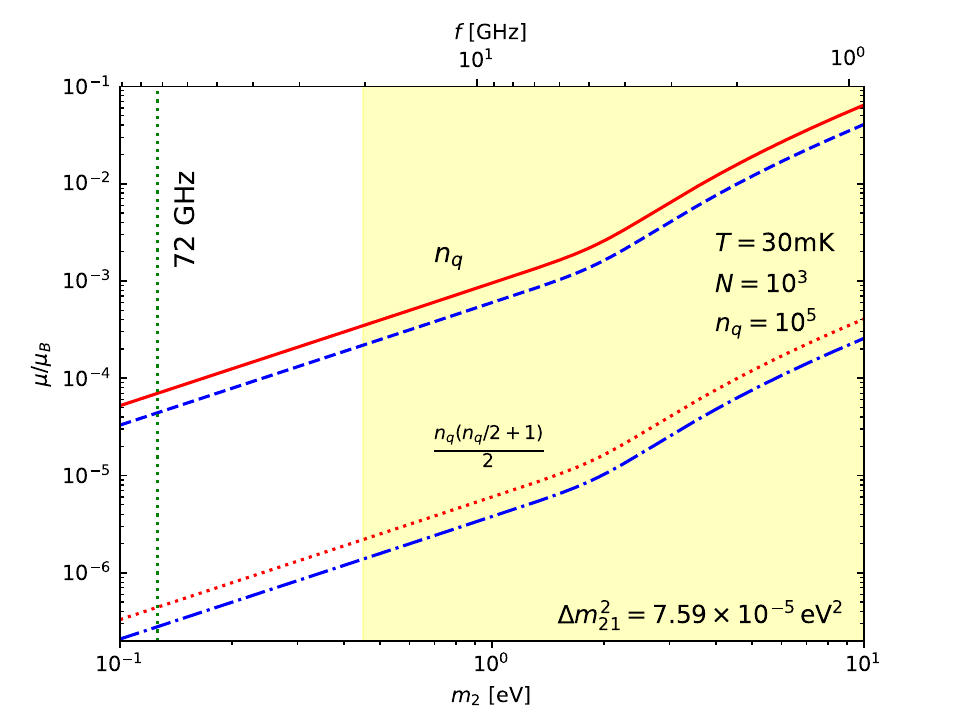} \hspace*{-0.5cm}
\includegraphics[width=0.5\textwidth]{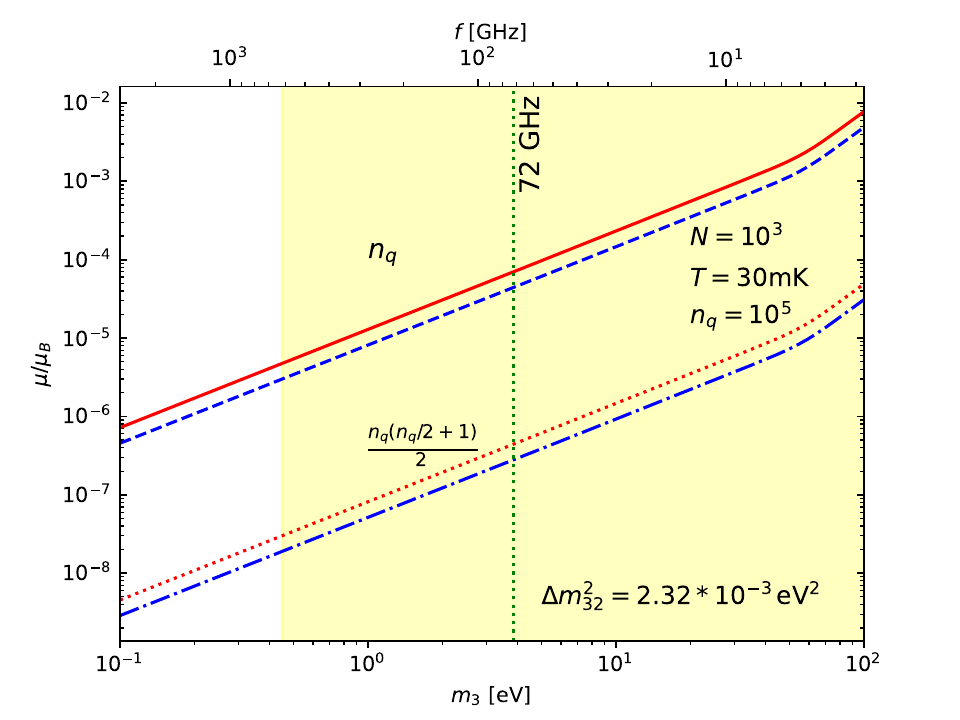}   
\caption{The 5$\sigma$ discovery (red, solid) and 2$\sigma$ exclusion (blue, dashed) significance of the transmon qubit device for the neutrino transition magnetic moment $\mu$ (in the $\mu_B$ unit) as a function of $m_i$ for $\Delta m^2_{21} = 7.59 \times 10^{-5} \ {\rm eV^2}$ (left) and $\Delta m^2_{32} = 2.32 \times 10^{-3} \ {\rm eV^2}$ (right), adopting $n_q= 10^5$ and $N=10^3$. 
We adopted a 14-second data collection for each frequency as per the measurement protocol with a varying coherence time $\tau = \dfrac{2\pi Q}{\omega}$ with $Q \approx 10^6$ following Ref.~\cite{Chen:2022quj}. 
We use a temperature of $T=30$~mK to estimate the dark-count background.
The (red, dotted) and (blue, dot-dashed) curves represent 5$\sigma$ discovery and 2$\sigma$ exclusion for Dicke states. 
The vertical green-dotted line marks the 72 GHz frequency of the high-frequency superconducting qubit investigated in Ref. \cite{Anferov:2024ljv}.
The yellow-shaded region ($m_\nu > 0.45$ eV) represents the mass range excluded at a 90\% confidence level by the KATRIN experiment \cite{KATRIN:2024cdt}, which provides the most robust, strictly model-independent upper bound on the neutrino mass derived from direct laboratory kinematics.}
\label{fig:CNBbounds}
\end{figure*}

We find that the transition probability $p_\ast$ for a single transmon qubit can be expressed as
\begin{widetext}
\begin{align}
p_\ast  \simeq &\, p_\ast^0 \times 
  \left( \frac{d}{100\ \mu{\rm m}} \right)^2
  \left( \frac{C}{0.1\ {\rm pF}} \right)  
  \left( \frac{f}{1\ {\rm GHz}} \right)
  \left( \frac{\tau}{100\ \mu{\rm s}} \right)^3
  \left ( \frac{\mu}{10^{-11}\ \mu_B} \right)^2
  \left ( \frac{\Delta m_{ij}^2}{10^{-5}\ \rm eV^2} \right)^4 
  \left ( \frac{
  \,n_\nu}{112\ \rm cm^{-3}} \right)
  \left ( \frac{\rm 0.05\ eV}{m_i} \right)^4
\label{eq:p_ge}
\end{align}
\end{widetext}
where the reference probability is $p_\ast^0=2.1 \times 10^{-27}$ and $f=\dfrac{\omega}{2\pi}$ is related to neutrino mass parameters as follows:
\begin{align}
  f \simeq 24 \ {\rm GHz} \times
  \left( \frac{\Delta m^2_{ij}}{10^{-5}\ {\rm eV^2}} \right) \left ( \frac{\rm 0.05\ eV}{m_i} \right ) \, .
  \label{eq:frequency}
\end{align}

The sensitivity for the neutrino magnetic moment $\mu$ is derived by comparing $N_{\rm sig}$ and $N_{\rm dark}$. We show in Fig. \ref{fig:CNBbounds} the 5$\sigma$ discovery (red, solid) and 2$\sigma$ exclusion (blue, dashed) using Eqs.~\eqref{eq:5sigma} and \eqref{eq:2sigma} for the neutrino transition magnetic moment $\mu$ (in the $\mu_B$ unit) as a function of the mass of the decaying neutrino ($m_i$ in eV) for $\Delta m^2 = 7.59 \times 10^{-5} \ {\rm eV^2}$ (left) and $\Delta m^2 = 2.32 \times 10^{-3} \ {\rm eV^2}$ (right). We adopted a 14-second data collection for each frequency as
per the measurement protocol with a varying coherence time 
$\tau = \dfrac{2\pi Q}{\omega}$ with $Q \approx 10^6$ following Ref.~\cite{Chen:2022quj}. 
The backgrounds are calculated using the same methodology described in the previous section. 
Current quantum devices lack the sensitivity required to detect the C$\nu$B, so we consider a futuristic scenario.
We adopt $n_q= 10^5$ as our (aggressive) benchmark, motivated by IBM's roadmap to build a 100,000-qubit quantum computer by 2033 through advances in error correction, modular design, and scalable engineering \cite{IBM}. We also considered $N=10^3$ units, 
each hosting an $n_q=10^5$ qubit-system. Our results show that the squared-mass difference of $\Delta m^2 \sim 10^{-5}\,  {\rm eV^2}$ may be probed with the frequencies of $1-10$~GHz, while $\Delta m^2 \sim 10^{-3} \,  {\rm eV^2}$ requires much higher frequency. 
The potential enhancements provided by Dicke states are illustrated by the red-dotted and blue-dot-dashed curves, which correspond to the 5$\sigma$ discovery and 2$\sigma$ exclusion limits, respectively.
\begin{table*}[th!]
\renewcommand\arraystretch{1.5}
\centering
\begin{tabular}{c||c|c|c|c|c}
\toprule
\hline \hline 
~\textbf{Ion Species}~ & ~\textbf{Transition Frequency ($f^\ast$)}~ & ~$\tau^\ast$ (sec)~ & ~$m_i^\ast$ (eV)~ & ~$m_{\rm ion}$ (amu) ~& ~$p^0_\ast$~ \\
\hline \hline 
\midrule
$^{171}\mathrm{Yb}^+$ \cite{olmschenk2007manipulation,Wang_Luan_Qiao_Um_Zhang_Wang_Yuan_Gu_Zhang_Kim_2021} & \SI{12.6}{\giga\hertz} (2.38 cm)            
 & 10  & 0.09595 & 170.9 & $8.59\times 10^{-26}$ \\
$^{43}\mathrm{Ca}^+$ \cite{ballance2016high,harty2014high}  &  \SI{3.2}{\giga\hertz} (9.4 cm)                 &50 & 0.3778 & 42.9588 & $7.0 \times 10^{-25}$ \\
\hline \hline 
\bottomrule
\end{tabular}
\caption{
The same as Table \ref{table:trappedionDM} but for the neutrino decays.}
\label{table:trappedion}
\end{table*}

Similarly, the transition probability $p_\ast$ for trapped ions can be calculated as 
\begin{widetext}
\begin{align}
p_\ast &=p_\ast^0 \times
\left ( \frac{\mu}{10^{-11} \mu_B} \right )^2
\left ( \frac{
n_\nu}{112 \, {\rm cm^{-3}}} \right )
\left ( \frac{\Delta m^2}{10^{-5} \,{\rm eV^2}} \right )^4 
\left ( \frac{m_i^\ast}{m_i} \right )^4
\left ( \frac{\tau}{\tau^\ast} \right )^3
\left ( \frac{f^\ast}{f} \right ) \, , 
\end{align}
\end{widetext}
where the resonant frequency is given by 
\begin{equation}
f \simeq f^\ast \times
  \left( \frac{\Delta m^2_{ij}}{10^{-5}\ {\rm eV^2}} \right) \left ( \frac{m_i^\ast}{m_i} \right ) \, .
\end{equation}
Table \ref{table:trappedion} summarizes trapped ion qubit parameters for two ion species, which show slightly higher transition probability compared to the transmon case.

\bigskip
As in the DM case, one can reinterpret the DarQ experiment results to constrain the neutrino transition magnetic moment.
Imposing the condition $\bar E_{(\rm eff)} < E_{\rm max}^{\rm DarQ}$ yields a bound of $\mu < 334 \, \mu_B$.
While this bound from DarQ is much weaker than the current limit $\mu < 10^{-11} \, \mu_B$ \cite{Giunti:2014ixa,Brdar:2020quo}, it is notable as the first constraint derived using a quantum device.

\section{Conclusions and discussion}
\label{sec:conclusion}

In this work, we have discussed a novel approach to probe the radiative decays of extremely weakly interacting particles, using radiative decays of dark matter and cosmic neutrinos as representative examples.
We have focused on two of the leading platforms: superconducting transmon qubits and trapped ion systems. By modeling the effective electric fields induced by decay photons, we have evaluated the response of these quantum sensors across two relevant particle physics scenarios.

Our results demonstrate that quantum devices can effectively explore new regions of parameter space previously inaccessible to traditional detectors. Specifically, we find that radiative decays of DM particles can be probed with current quantum technologies. In contrast, probing neutrino magnetic moments beyond existing experimental bounds will require further advancements in coherence time and scalability. For a roadmap toward building large-scale quantum devices, we refer the reader to Refs. \cite{Dwave,IBM,Fujitsu,Google,IonQ}. 

Looking ahead, several directions offer promising avenues for improvement. Enhancing coherence times---particularly through novel superconducting cavity architectures \cite{Huang_2020,Krantz_2019} or advanced trapped ion systems \cite{Bruzewicz_2019}---can substantially boost sensitivity, as the signal strength scales favorably with the coherence time, i.e., $p_\ast \sim \tau^3$.
Superconducting quantum memories currently exhibit coherence times on the order of a few milliseconds, and extending beyond this limit has remained a significant challenge. Ref.~\cite{Milul:2023nbc} has demonstrated a single-photon qubit encoded in a novel superconducting cavity, achieving a coherence time of 34 milliseconds. In Ref.~\cite{Anferov:2024ljv}, by developing a 72 GHz millimeter-wave superconducting qubit that operates at 0.87K using simpler refrigeration, researchers have eliminated the need for rare cooling isotopes while opening up an unexplored, low-noise energy range for advanced quantum computing and sensing.
 While higher operating frequencies in superconducting transmon qubits offer potential benefits such as faster gate operations and more efficient control, they also present challenges. Primarily, higher frequencies exacerbate decoherence due to increased sensitivity to environmental noise and quasiparticle effects, potentially limiting coherence times. Furthermore, the development and integration of hardware capable of supporting these higher frequencies can be complex. 

The development of multi-qubit entanglement protocols and collective interaction schemes amplifies detection prospects while suppressing background noise. Furthermore, expanding the search to include additional interactions, such as millicharged or anapole couplings, and exploring sterile neutrino decay channels, may broaden the scope of potential discoveries.

Although practical implementation remains challenging, continued progress in quantum hardware and control techniques may soon bring these ideas within experimental reach. 
Quantum sensing thus holds considerable promise as a new frontier in fundamental physics, offering a pathway to explore elusive phenomena like the C$\nu$B and radiative decays of DM candidates.

Finally, we note that the interpretation of discovery with quantum devices requires careful analysis, since there are many possible physics scenarios that could induce the effective electric field.

\section*{Acknowledgements}
We thank Takeo Moroi, Kaladi Babu, André de Gouvêa, Vedran Brdar, Kyu Jung Bae, and Talal Chowdhury for helpful discussions. 
The authors wish to acknowledge the Center for Theoretical Underground Physics and Related Areas (CETUP), the Institute for Underground Science at Sanford Underground Research Facility (SURF), and the South Dakota Science and Technology Authority for hospitality and financial support, as well as for providing a stimulating environment. 
The work of ZD and MA is supported in part by the US DOE under Award No DE-SC0024673.
The work of DK is supported in part by the NSF Grant No. PHY-2609913.
The work of KK is supported in part by the  NSF Grant No. PHY-2609759. 
The work of MP is supported by the National Research Foundation of Korea (NRF) grant funded by the Korea government (MSIT) (No.RS-2025-00564488). 


\appendix

\section{Photon residence time and cavity losses} \label{sec:photonresidence}

The survival of a decay photon inside the cavity is limited by wall losses, which determine the effective photon residence time. To estimate this timescale, we first compute the average path length between wall reflections. For a propagating ray inside a convex cavity, the mean chord length theorem gives the average distance between successive wall collisions as \cite{Chandrasekhar1960}

\begin{equation}
\langle \ell \rangle = \frac{4V}{A},
\label{eq:mean_chord}
\end{equation}
where $V$ is the cavity volume and $A$ its surface area. For a spherical cavity of radius $R_c$, this yields
\begin{equation}
\langle \ell \rangle = \frac{4R_c}{3}.
\label{eq:sphere_chord}
\end{equation}
The corresponding mean time between reflections is
\begin{equation}
t_{\rm bounce} = \frac{\langle \ell \rangle}{c}.
\label{eq:bounce_time}
\end{equation}
If the wall power reflectivity at the relevant microwave frequency is $R$, then after $N$ reflections the photon intensity scales as $I_N = I_0 R^N$. Since reflections occur at intervals $t_{\rm bounce}$, the intensity decays exponentially,
\begin{equation}
I(t) = I_0 R^{t/t_{\rm bounce}}
     = I_0 e^{-t/\tau_{\rm cav}},
\label{eq:intensity_decay}
\end{equation}
with photon ``lifetime''
\begin{equation}
\tau_{\rm cav} = \frac{t_{\rm bounce}}{-\ln R}.
\label{eq:cavity_lifetime_exact}
\end{equation}

In the high-reflectivity limit $R \simeq 1$, one may use $-\ln R \simeq 1 - R$, giving
\begin{equation}
\tau_{\rm cav} \simeq \frac{t_{\rm bounce}}{1 - R}
                   = \frac{\langle \ell \rangle}{c(1 - R)}.
\label{eq:cavity_lifetime_approx}
\end{equation}
This expression corresponds to the standard cavity ring-down result \cite{Siegman1986,Berden2000}.
For a meter-scale cavity with $R_c = 0.5\,{\rm m}$, one finds $\langle \ell \rangle \simeq 0.67\,{\rm m}$ and $t_{\rm bounce} \simeq 2.2\,{\rm ns}$. The resulting photon lifetime is
\begin{equation}
\tau_{\rm cav} \approx
\begin{cases}
0.22\,\mu{\rm s}, & R = 0.99, \\
2.2\,\mu{\rm s}, & R = 0.999, \\
22\,\mu{\rm s}, & R = 0.9999.
\end{cases}
\label{eq:lifetime_numeric}
\end{equation}
Thus, achieving photon survival on $\mu{\rm s}$ timescales requires a reflectivity of at least $R \gtrsim 0.999$.
For resonant microwave cavities, the photon lifetime may equivalently be expressed in terms of the quality factor $Q_{\rm cav}$,
\begin{equation}
\tau_{\rm cav} = \frac{2\pi Q_{\rm cav}}{\omega},
\label{eq:Q_relation}
\end{equation}
where $\omega$ is the qubit resonance frequency \cite{Pozar2011}. The geometric estimate above provides an order-of-magnitude lifetime appropriate for multi-reflection photon trapping; a more precise determination would compute the conductor-loss-limited quality factor of the cavity at the operating frequency.

In our sensitivity estimates, the effective photon residence time available to excite a qubit is therefore given by 
\begin{equation}
t_{\rm eff} = \min(\tau_{\rm cav},\, \tau_{\rm qubit}),
\label{eq:effective_time}
\end{equation}
ensuring that photon buildup is properly limited by either cavity losses or qubit coherence time.

\bibliography{refs}

\end{document}